\documentclass{PoS}
\usepackage{subfigure}
\usepackage{wrapfig}

\title{The Suzaku Hard X-ray Survey on the Galactic Center Region}

\ShortTitle{The Galactic Center region}

\author{\speaker{Takeshi~Go~Tsuru}$^{\ 1}$, 
	H.~Uchiyama, M.~Nobukawa, M.~Sawada, 
	S.G.~Ryu, R.~Fukuoka, K.~Koyama\\
        Department of Physics, Faculty of Science, Kyoto University, 
        Kitashirakawa, Sakyo, Kyoto 606-8502, JAPAN\\
        $^1$E-mail: \email{tsuru@cr.scphys.kyoto-u.ac.jp}}

\abstract{
Diffuse X-rays from the Galactic center (GC) region were found to exhibit
many K-shell lines from iron and nickel atoms in the 6--9 keV band.
The strong emission lines seen in the spectrum 
are neutral iron K$\alpha$ at 6.4~keV, He-like iron K$\alpha$ at 6.7~keV, 
H-like iron Ly$\alpha$ at 6.9~keV, and He-like iron K$\beta$ at 7.8~keV. 
Among them, the 6.4~keV emission line is a probe of non-thermal phenomena. 
We have detected strong 6.4~keV emission in several giant molecular clouds, 
some of which were newly discovered by Suzaku. 
All the spectra exhibit large equivalent widths of 1-2~keV and absorption columns of 
$2-10\times 10^{23}\ {\rm H\ cm}^{-2}$. 
We found time variability of diffuse 6.4~keV emission in the Sgr B2 region comparing 
the maps and spectra obtained from 1994 to 2005 with ASCA, Chandra, XMM-Newton and Suzaku. 
We also report discovery of K$\alpha$ lines of neutral argon, calcium, chrome, and 
manganese atoms in the Sgr~A region. 
We show that the equivalent width of the 6.4~keV emission line detected in X-ray faint region 
against the 6.4 keV-associated continuum (power-law component) is $\sim 800\ {\rm eV}$. 
These features are naturally explained by the X-ray reflection nebula scenario rather 
than the low energy cosmic-ray electrons scenario. 
On the other hand, a 6.4~keV clump, G~0.162$-$0.217, 
discovered at the south end of the Radio Arc has 
a small equivalent width of 6.4~keV emission line of $\sim200\ {\rm eV}$. 
The Radio Arc is a site of relativistic electrons. 
Thus, it is conceivable that the X-rays of G~0.162$-$0.217 are due 
to low energy cosmic-ray electrons 
}

\FullConference{The Extreme sky: Sampling the Universe above 10 keV -
  extremesky2009,\\ October 13-17, 2009\\ 
  Otranto (Lecce) Italy}

\begin{document}
\section{Introduction}
\begin{figure}
  \centering
  \includegraphics[width=.8\textwidth]{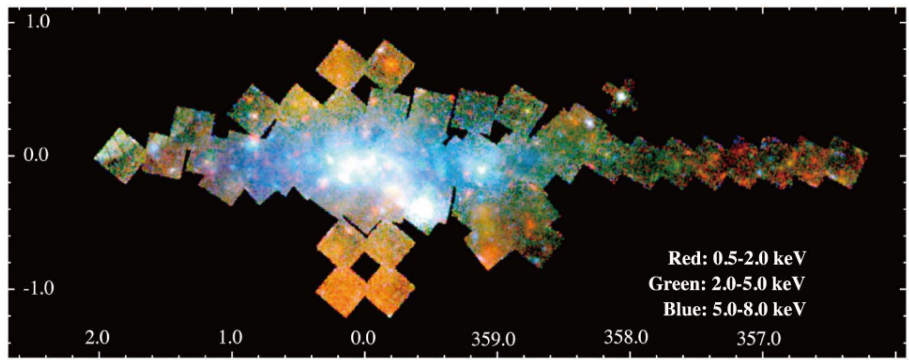}
  \caption{The X-ray image of the Galactic center region obtained with the Suzaku XIS. 
    The colors indicate X-ray energy bands - red (0.5--2.0~keV), 
    green (2.0--5.0~keV), and blue (5.0--8.0~keV). }
  \label{GC60FOV_3bandImage}
\end{figure}

The X-ray Imaging Spectrometer (XIS) on board the Suzaku satellite is 
the X-ray CCD camera system which has the large effective area, 
the low and the stable non X-ray background in the 0.3--12~keV band. 
The diffuse hard X-ray emission from the Galactic center region, 
particularly in the iron K-shell band, is the best target for Suzaku. 
We have been making survey observation as a Suzaku key project. 
We have finished about 60 pointings with a total exposure 
time of $\sim 3000$~ksec and 
published 22 papers so far (Figure~\ref{GC60FOV_3bandImage}). 
Figure~2 shows an excellent spectrum of the Sgr A region 
\cite{KoyamaPASJ07_GCplasma_Fe_Ni_Suzaku}.
The results from the key project can make a Suzaku legacy in the X-ray astronomy. 

\begin{figure}
  \centering
  \includegraphics[width=.6\textwidth]{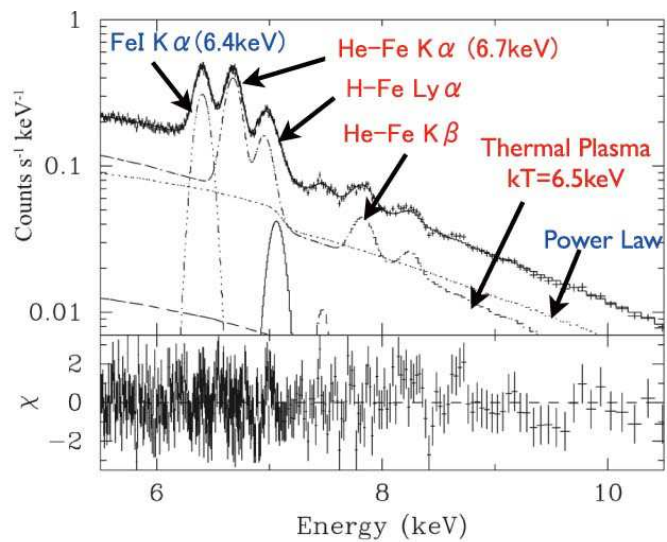}
  \caption{The averaged spectrum of the Sgr~A region in the 5.5-11.5~keV band 
    taken with the Suzaku XIS \cite{KoyamaPASJ07_GCplasma_Fe_Ni_Suzaku}. 
    The spectrum is fitted with a model of a 
    collisional ionization equilibrium plasma plus a power-law with 
    three Gaussian lines and an iron absorption edge.}
  \label{SgrA_Spec_Koyama07}
\end{figure}

The spectrum of  the diffuse X-rays from the Galactic center region exhibits
many K-shell lines from iron and nickel atoms in the 6--9 keV band 
(Figure~\ref{SgrA_Spec_Koyama07}).
The strong emission lines seen in the spectrum 
are neutral iron K$\alpha$ at 6.4~keV, He-like iron K$\alpha$ at 6.7~keV, 
H-like iron Ly$\alpha$ at 6.9~keV and He-like iron K$\beta$ at 7.8~keV.
We determined the electron temperature from the flux ratio of 
K$\beta$ and K$\alpha$ emission lines from He-like 
iron~\cite{KoyamaPASJ07_GCplasma_Fe_Ni_Suzaku}. 
The ionization temperature was obtained from the fluxes of 
the emission lines of H-like iron Ly$\alpha$ and He-like iron K$\alpha$. 
Both temperatures are $kT\sim6.5$~keV. 
Thus, we can safely fix the continuum shape of the thermal component 
at the temperature of $kT=6.5$~keV.
An additional power-law component with the photon index $\Gamma=1.4^{+0.5}_{-0.7}$ 
is necessary to fit the observed continuum spectrum~\cite{KoyamaPASJ07_GCplasma_Fe_Ni_Suzaku}. 
The power-law component may have a non-thermal origin and will 
dominate in the X-ray spectrum above 10~keV. 
The spectral index is similar to those of X-ray spectra in the 6.4 keV clouds.
It indicates that the 6.4~keV emission line is related to the power-law component and 
can be a probe of the non-thermal phenomena. 

\section{The 6.4~keV line emission from Giant Molecular Clouds}
\begin{figure}
  \centering
  \includegraphics[width=.8\textwidth]{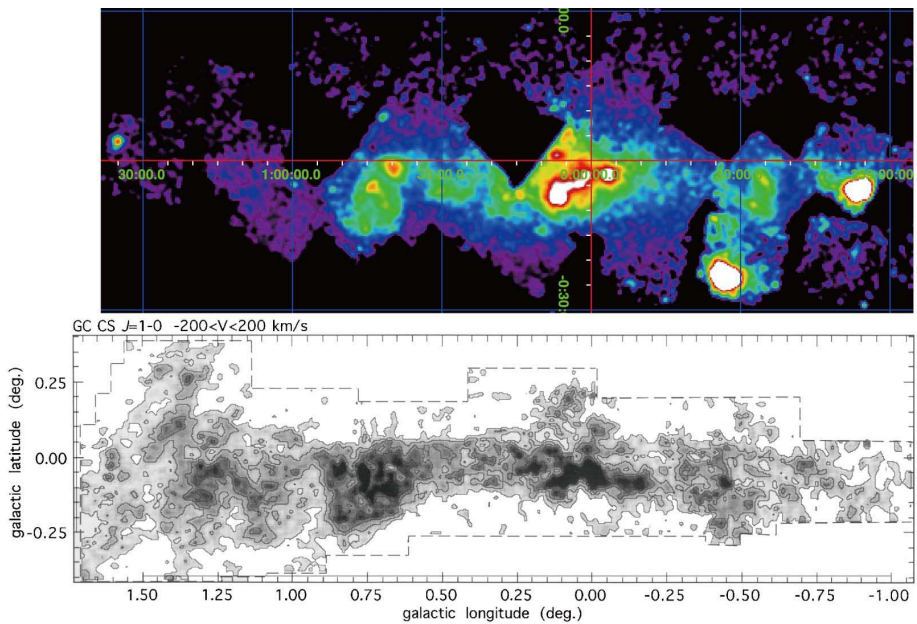}
  \caption{The maps of the 6.4~keV line band (top) and the velocity-integrated 
    carbon monosulfide line (bottom)~\cite{TsuboiApJS99_GC_CS_Survey}. }
  \label{fig3}
\end{figure}

Figure~\ref{fig3} shows the maps of the 6.4~keV line and 
the carbon monosulfide line which traces molecular clouds\cite{TsuboiApJS99_GC_CS_Survey}. 
The 6.4~keV emission shows a clumpy distribution while 
the 6.7~keV emission has few distinctive structures. 
Figure~\ref{fig3} indicates that the 6.4keV line emission generally 
traces the distribution of the carbon monosulfide line emission. 
It suggests that the iron K-shell fluorescence line emission 
occurs in the molecular clouds.

Two models have been proposed for the origin of the 6.4~keV emission line. 
One is the X-ray photo-ionization by external X-ray sources 
(the XRN scenario)~\cite{KoyamaPASJ96_GC_ASCA, SunyaevMN98_GC_XRN_XrayFluxHist, ParkApJ04_GC20pc_DiffuseLine}. 
Since no irradiating source capable of powering the 6.4~keV line is found, 
Koyama et~al. (1996) proposed a scenario of a past X-ray outburst of the super-massive
black hole at Sgr A$^*$.
The other is the inner-shell ionization by the impact of 
low energy cosmic-ray electrons 
(the LECRe scenario)~\cite{2002Yusef-ZadehApJL_MC_6.4keV_LECRe}. 

The power-law component is the Thomson scattered radiation of the incident X-rays 
in the XRN scenario, or the bremsstrahlung in the LECRe one. 
The expected X-ray spectrum is different between the two scenarios. 
The 6.4~keV line is expected to have a larger equivalent width in the X-ray reflection 
($\sim 1~{\rm keV}$) than that in the LECRe ($\sim 0.3~{\rm keV}$). 
Hard X-rays photo-ionizing iron can reach a deep portion of a molecular cloud, 
but electrons easily stop at the surface due to the ionization loss. 
Thus, larger absorption column is expected in the XRN model
than in the LECRe one. 

We have detected several 6.4~keV clumps so far, 
some of which were newly discovered by 
Suzaku~\cite{2009NakajimaPASJ_SgrC_XRN, KoyamaPASJ07_SgrB2_NewDiffuseSource_Suzaku, 2008NobukawaPASJ_SgrB1_SNR_XRN, 2009FukuokaPASJ_SouthEnd_RadioArc_Suzaku}. 
All the spectra have large equivalent widths of 1--2~keV and 
absorption columns of $2-10\times 10^{23}~{\rm H\ cm^{-2}}$. 
These features are naturally explained by the XRN scenario rather than 
the LECRe one.

\begin{figure}
  \centering
  \includegraphics[width=.9\textwidth]{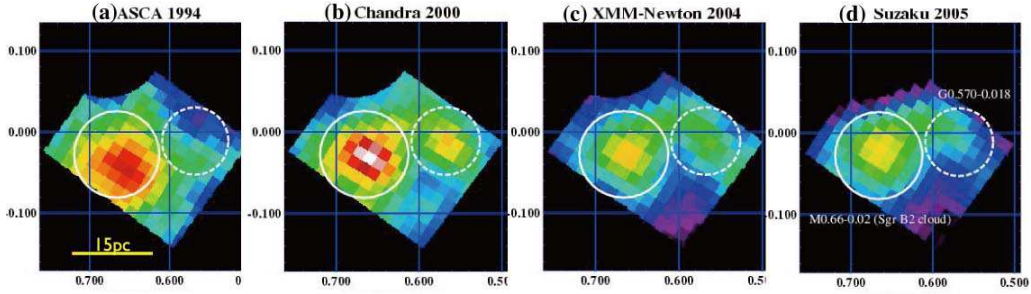}
  \caption{Surface brightness maps at the energy band of 6.0--7.0~keV including 
  iron K lines obtained with (a) ASCA SIS in 1994, (b) Chandra ACIS-I in 2000, 
  (c) XMM-Newton MOS and PN in 2004, and 
  (d) Suzaku XIS in 2005~\cite{2009InuitPASJ_SgrB2_TimeVari}. 
	Pixel size is $50"\times 50"$ in each case.}
  \label{fig4}
\end{figure}

We found time variability of the diffuse 6.4~keV emission in the Sgr~B2 
region~\cite{2008KoyamaPASJ_SgrB_TimeVariabiliry, 2009InuitPASJ_SgrB2_TimeVari}. 
Figure~\ref{fig4} shows the maps of the line components in 6--7~keV from 1994 to 2005. 
The two region marked with solid and dashed circles in Figure~\ref{fig4} became 
bright in 2000 but faded in 2004 and 2005. 
We analyzed the X-ray spectra of the observations, and found that the flux of 
the 6.7~keV emission line had been constant through the decade observations. 
On the other hand, the flux of the 6.4~keV emission line changed by a factor of two. 

The decay time of the flux is about 10~yrs. 
The size of the 6.4~keV clump resolved by Chandra is 10~lyrs. 
Electrons with the energy range of 10--100~keV, 
where the cross section of inner shell ionization of iron is maximum, 
are unable to travel 10~lyrs in 10~yrs. 
On the other hand, the XRN model explains this time variability. 

\section{Discovery of K$\alpha$ lines from Neutral Atoms}
\begin{figure}
  \centering
  \includegraphics[width=0.92\textwidth]{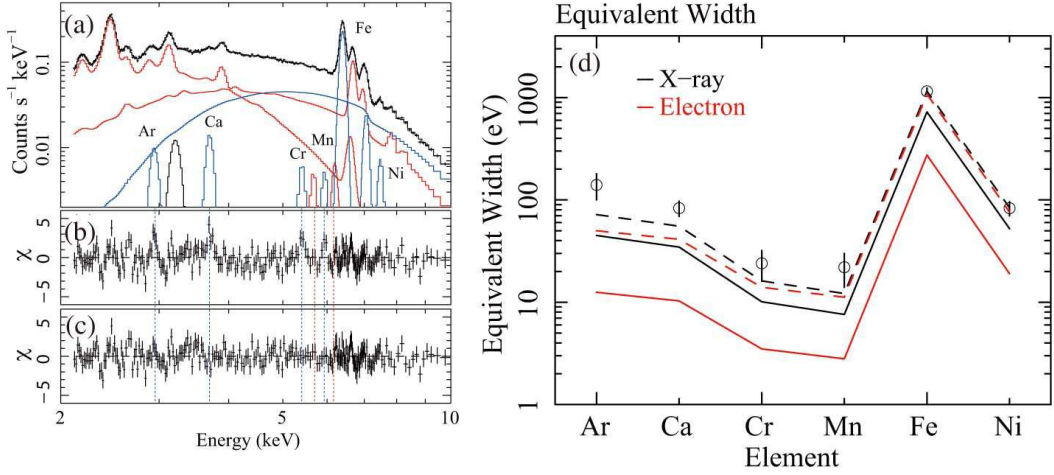}
  \caption{(a) Suzaku XIS-FI spectrum of the bright neutral clumps  in the Sgr~A region~\cite{2010NobukawaPASJ}. 
	The figure is also plotted with the best fitted model consisting of 2-$kT$ APEC, 
	He-like Cr and Mn K$\alpha$ lines (red),  a power-law, neutral Ar, Ca, Cr, Mn, Fe, Ni K$\alpha$ 
	and iron K$\beta$ emission lines (blue). 
	(b) Residual of the fitting with a model of 2-kT APEC , He-like Cr and Mn K$\alpha$ lines 
	a power-law, and neutral lines of Fe I K$\alpha$, K$\beta$, Ni I K$\alpha$. 
	(c) Same as (b) but the neutral Ar, Ca, Cr, and Mn K$\alpha$ lines are added. 
	(d) Equivalent widths of K$\alpha$ line of various neutral atoms. 
	Black and red lines are the calculated value for the X-ray (XRN) and 
	electron (LECRe) scenarios, respectively.
	The data points marked with the open circles are observed value in our work. 
	Errors are estimated at the 90\% confidence level. 
	The black and red dashed lines are to guide eyes, 
	which are XRN and LECRe scenarios in 1.6 solar and 4.0 solar abundances, respectively.}
  \label{fig5}
\end{figure}
%
We have shown the results from the neutral iron K$\alpha$ emission line (6.4~keV). 
How about the other elements ? 
These would provide new information to constrain the origin of 
the neutral clumps in the GC region. 
Since the relative existences of the lighter elements are 10 to 100 times smaller 
than that of iron, we search the brightest 6.4~keV clump located in the Sgr~A region 
for their fluorescence lines\cite{2010NobukawaPASJ}. 
Figure~\ref{fig5}a shows the spectrum of the clump. 
We made a spectral fitting with a model of two temperature thermal plasma components 
plus a power-law, three Gaussians for the neutral iron K$\alpha$, K$\beta$, and 
nickel K~$\alpha$ emission lines. 
There remained residuals at the energies of K emission lines 
of the highly ionized chrome and manganese. 
Since the APEC model, the thermal plasma code used in the fitting, 
dose not contain the chrome and manganese K-lines, 
we added Gaussians for the lines. 
These lines were firstly detected from the Galactic center region.
The temperatures of the plasmas are 1~keV and 7~keV. 
The absorption column density for the plasma components is $7\times 10^{22}~{\rm H\ cm}^{-2}$, 
which is typical for the sources in the Galactic center region.
Abundances of the elements in the plasmas are $\sim 1.9$ solar for sulfur, argon, and calcium, 
$\sim 1.2$ solar for iron and $\sim 1.6$ for nickel. 
The power-law component has the index of $\sim 2$ and requires 
absorption column density of $1.7\times 10^{23}~{\rm H\ cm}^{-2}$ which is significantly 
larger than that for the plasma components. 
There still remained four line-like residuals 
whose energies correspond to the neutral argon, calcium, chrome, and manganese 
K$\alpha$ lines (Figure~\ref{fig5}b).  
So, we added Gaussians modeling them to the spectral model and 
found that it improved the fitting significantly (Figure~\ref{fig5}c). 
We discovered the K-shell lines of neutral argon, calcium, chrome, manganese 
from the bright neutral clump toward the Sgr~A. 


Figure~\ref{fig5}d shows the equivalent widths of the neutral lines 
against the power-law component. 
In order to investigate the origin of the neutral lines, 
we calculated the equivalent widths in the two scenarios of 
the XRN and the LECRe using Geant~4. 
In this calculation, we assumed that density of the molecular cloud is uniform, 
the absorption column is $10^{23}~{\rm H\ cm}^{-2}$, and elemental abundances are solar.
For the LECRe scenario, the metal abundances in the molecular cloud must be $\sim 4$ 
times larger than the solar value. 
The XRN scenario requires $\sim 1.6$ solar abundances. 
Since the molecular cloud may be formed by condensation
of the ambient materials, the abundances should be similar to those in the plasma
component in the GC region. 
The plasma components have the abundances of $\sim 1.2-1.9$, 
which agrees well with those required in the XRN scenario. 
Thus, we suggest that the K$\alpha$ emission lines from neutral elements 
are due to the X-ray reflection. 


\section{The 6.4~keV emission lines in X-ray faint regions}
\begin{figure}
	\centering
	\includegraphics[width=0.92\textwidth]{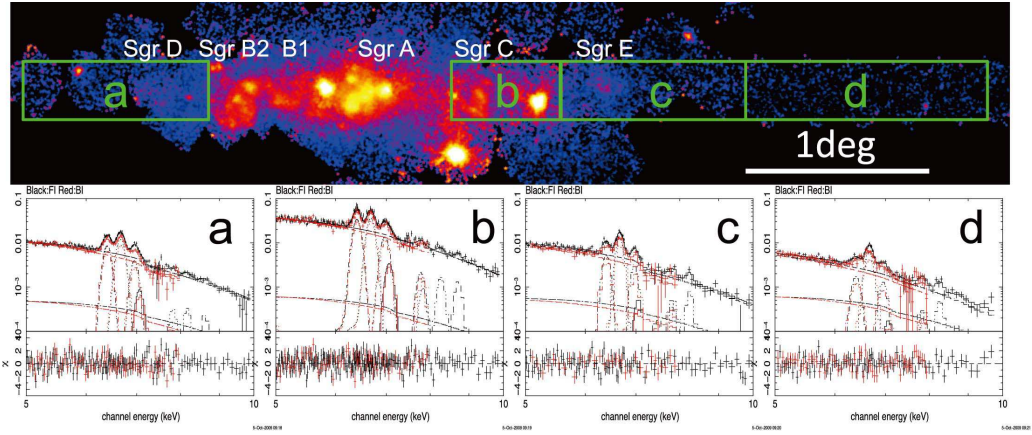}
	\caption{The spectra of the X-ray faint regions (bottom). 
	The spectral extraction regions are overlaid on the 6.4~keV map 
	(top)~\cite{2010UchiyamaPhD}.}
	\label{fig6}
\end{figure}

\begin{wrapfigure}{r}{0.4\textwidth}
	\vspace{-10mm}
	\centering
	\includegraphics[width=0.40\textwidth]{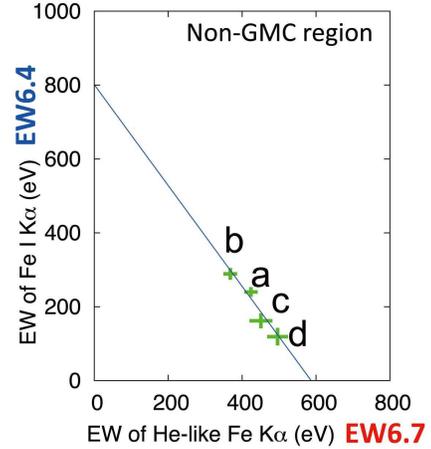}
	\caption{The relation between the equivalent widths of the 6.4~keV and 6.7~keV emission lines 
	in the spectra of the X-ray faint regions (Figure~6)~\cite{2010UchiyamaPhD}.}
	\label{fig7}
	\vspace{-5mm}
\end{wrapfigure}
%
So far, we have shown the results from the X-ray bright clumps 
as the Sgr~A and B2 regions. 
We now focus on X-ray faint region in this section. 
Figure~\ref{fig6} shows the regions investigated and their spectra. 
Significant 6.4~keV emission lines are seen as those in the bright regions. 
Since the emission is so diffuse that the selection of the background region is difficult, 
we made a different approach. 



Koyama et~al. (2009) and Nakajima et~al. (2009) showed that 
the Galactic center diffuse X-rays (GCDX) in the Sgr~A and Sgr~C
regions is phenomenologically decomposed 
into the 6.7-keV line plus an associated continuum (6.7-component) 
and the 6.4-keV line plus an associated continuum (6.4-component). 
The equivalent widths (EWs) of the 6.4~keV ($EW_{6.4}$) and the 6.7~keV ($EW_{6.7}$) lines 
for the Sgr~A region are given by the relation 
$EW_{6.7} + 0.5(\pm 0.06)\times EW_{6.4} = 
0.62(\pm 0.07)$~keV~\cite{2009Koyama_PASJ_GC_Plasma_FeI, 2009NakajimaPASJ_SgrC_XRN}. 
In the limit of $EW_{6.7}\rightarrow 0$, 
$EW_{6.4}$ in the 6.4-component was successfully estimated to be $1.2\pm 0.2$~keV. 

We used the same method to decompose the GCDX in the X-ray faint 
regions~\cite{2010UchiyamaPhD}. 
Figure~\ref{fig7} shows the correlation plot between the observed EWs. 
This relation indicates that $EW_{6.4}$ in the 6.4-component is estimated to be 
$\sim 800$~eV at $EW_{6.7}\rightarrow 0$. 
The values is significantly lower than the one expected in the XRN scenario but 
higher than that of the LECRe model. 
Further investigation is necessary. 


%

\section{G~0.162$-$0.127: a 6.4 keV clump due to LECRe}
\begin{figure}
  \centering
  \includegraphics[width=.8\textwidth]{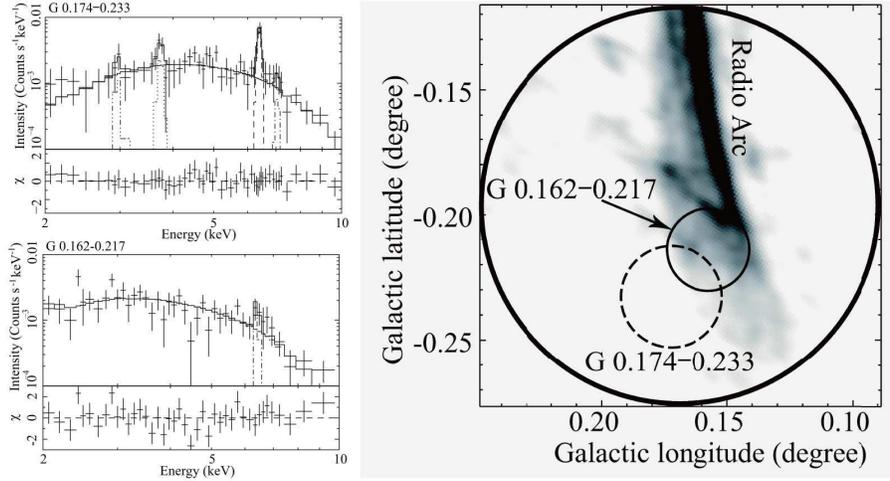}
  \caption{(left) Suzaku XIS (FI) spectra of G~0.174-233 and G~0.162-0.217 
  with the best-fit models of a power-law and Gaussian(s)~\cite{2009FukuokaPASJ_SouthEnd_RadioArc_Suzaku}.
  The three emission lines seen in G~0.174-233 are neutral calcium, argon K$\alpha$,  iron K$\alpha$ and K$\beta$. 
  The Gaussian shown in G~0.162-0.217 is neutral iron K$\alpha$. 
  (right) A 4.735 GHz radio continuum map of the south end of the Radio Arc 
  from the Very Large Array archive survey image. 
  The Very Large Array field is shown by the thick solid circle. 
  The positions of G~0.174-233 and G~0.162-217 are indicated 
  by the dashed and thin-solid circles.}
  \label{fig8}
\end{figure}

Among the radio non-thermal filaments discovered in the Galactic center region~\cite{LaRosaAJ00_GC_VLA90cm_WideField}, 
the most prominent one is the Radio Arc~\cite{1984YusefNature_RadioArc}. 
We found two 6.4~keV clumps, G~$0.174-0.233$ and G~$0.162-0.217$, 
around the south end of the Radio arc~\cite{2009FukuokaPASJ_SouthEnd_RadioArc_Suzaku}.
The right panel of Figure~\ref{fig8} shows their positions with the close-up view of the Radio arc. 
The spectrum of G~$0.174-0.233$ has prominent K$\alpha$, emission lines of 
neutral iron and calcium (the left top panel of Figure~\ref{fig8}). 
The EW of the iron 6.4~keV emission line is $\sim$950~eV. 
Thus, the XRN scenario is favored for the origin of G~$0.174-0.233$. 

The the spectrum of G~$0.162-0.217$ has significant 6.4~keV emission line 
(the left bottom panel of Figure~\ref{fig8}). 
The EW of the 6.4~keV line of $\sim 200$~eV is $\sim 20$\% of
the one expected by the XRN scenario but is consistent with the LECRe model ($\sim 300$~eV). 
The right panel of Figure~\ref{fig8} indicates that G~$0.162-0.217$ is located in the Radio Arc. 
Since the Radio Arc is a site of relativistic electrons, it may also include LECRe.
Thus, it is quite conceivable that the X-rays of G~$0.162-0.217$ are due to the LECRe. 
Assuming the electron energy of 10-100~keV, thick condition for a target cloud 
and the radiation yield of $\sim 4\times 10^{-5}$ due to bremsstrahlung, 
we obtained the electron energy density of $\sim 25~{\rm eV\ cm}^{-3}$. 
The magnetic field is estimated to be $\sim 30\mu{\rm G}$ 
in the condition of the equipartition. 


\end{document}